\documentclass[twocolumn]{aastex6}
\usepackage{natbib}
\usepackage{verbatim}
\usepackage{graphicx}
\usepackage[space]{grffile}
\usepackage{latexsym}
\usepackage{amsfonts,amsmath,amssymb}
\usepackage{url}
\usepackage[utf8]{inputenc}
\usepackage{fancyref}
\usepackage {cases}
\usepackage{float} 

\usepackage{ulem} 
\usepackage{color} 

\begin{document}

\title{Carbon Detonation Initiation in Turbulent Electron-Degenerate Matter}

\author{Robert Fisher$^{1, 2}$ } 
\email{robert.fisher@umassd.edu}
\author{Pritom Mozumdar$^1$}
\author{Gabriel Casabona$^1$}
 \affiliation {$^1$Department of Physics, University of Massachusetts Dartmouth, 285 Old Westport Road, North Dartmouth, MA 02740, USA}
\affiliation {$^2$Institute for Theory and Computation, Harvard-Smithsonian Center for Astrophysics, 60 Garden Street, Cambridge, MA 02138, USA}
\date{\today}


\begin{abstract}

Type Ia supernovae (SNe Ia) play a critical role in astrophysics, yet their origin remains mysterious. A crucial physical mechanism in any SN Ia model  is the initiation of the detonation front which ultimately unbinds the white dwarf progenitor and leads to the SN Ia.  We demonstrate, for the first time, how a carbon detonation may arise in a realistic three-dimensional turbulent electron-degenerate flow, in a new mechanism we refer to as turbulently-driven detonation.  Using both analytic estimates and three-dimensional numerical simulations, we show that strong turbulence in the distributed burning regime gives rise to intermittent turbulent dissipation which locally enhances the nuclear burning rate by orders of magnitude above the mean. This turbulent enhancement to the nuclear burning rate leads in turn to supersonic burning and a detonation front. As a result, turbulence plays a key role in preconditioning the carbon-oxygen fuel for a detonation. The turbulently-driven detonation initiation mechanism leads to a wider range of conditions for the onset of carbon detonation than previously thought possible, with important ramifications for SNe Ia models.

\end{abstract}



\maketitle

\section {Introduction}Type Ia supernovae (SNe Ia) are common luminous transients whose standardizable light curves play a crucial role in cosmology \citep {phillips93}. Recent observations of the nearby SN Ia 2011fe and SN 2014J have revealed SNe Ia arise in stellar systems with at least one carbon-oxygen white dwarf (WD) undergoing explosive nuclear burning \citep {nugentetal11, churazovetal14}. However, no stellar progenitor of a SN Ia event has ever been directly observed, and both their stellar progenitors and the mechanism of the explosion remains a subject of active investigation.


A crucial physical ingredient of SNe Ia is the detonation initiation within the carbon-oxygen WD. A detonation can be initiated in one of two ways: either directly, such as through a shock front, or spontaneously, through a preconditioned mix of heated fuel. 
Research in the SNe Ia context has focused upon spontaneous initiation, which has been explored by numerous authors, beginning with the work of Blinnikov and Khokhlov \citep {blinnikovkhoklhlov86, blinnikovkhoklov87}. This body of  spontaneous initiation work rests upon a theoretical foundation first developed by Zel'dovich and coworkers \citep {zeldovichetal70}, which invokes a gradient in induction timescales across a region, and which is referred to as the Zel'dovich gradient mechanism.

In this paper, we investigate the role which turbulence plays in preconditioning the thermodynamic state of  electron-degenerate matter of WDs under realistic conditions typical of leading SNe Ia channels. We focus upon carbon-oxygen fuel, though the turbulently-driven mechanism we describe relies only upon the fundamental physics of turbulence in an electron-degenerate nuclear reactive medium. Consequently, the basic detonation initiation mechanism is applicable to  turbulent regions of mixed carbon, oxygen, and helium electron-degenerate material, which are believed to arise in some SNe I scenarios \citep {guillochonetal10, pakmor_etal_2013}. We first briefly review the physics of the Zel'dovich gradient mechanism.

\section {Zel'dovich Gradient Mechanism} In its simplest form, the Zel'dovich gradient mechanism  begins with a subsonic laminar deflagration which is initiated at the peak of a temperature gradient. The deflagration proceeds to accelerate into a shock as it propagates down the temperature ramp. Provided that the temperature gradient is sufficiently shallow, the shock does not propagate rapidly in advance of the burning region, and the result is a detonation. In particular, a body of work based upon one-dimensional simulations \citep {arnettlivne94, khokhlovetal97, niemeyerwoosley97, ropkeetal07a, seitenzahletal09a, holcombetal13} establishes a critical length scale over which a detonation may be initiated, as a function of background density, temperature, and composition.  The Zel'dovich gradient mechanism generally requires a hot background temperature and a shallow temperature gradient on scales that are unresolved in global, full-star 3D SNe Ia simulations, making it challenging to connect the theory to global SNe Ia simulations.

We critically examine the tacit assumption underlying the Zel'dovich gradient mechanism; specifically, that the detonation initiation arises through the development of a laminar flame on a static background. In contrast to this underlying assumption of laminarity, all major SNe Ia channels predict the prevalent hydrodynamic flow conditions to be highly turbulent. For example, in the rapid dynamical phase of a WD merger, the secondary WD is tidally disrupted, and drives turbulence onto the primary through  strong shear flows. Turbulent energy is transported from large scales to small scales through the turbulent cascade, dissipating into heat energy, and mixed into deeper layers of the primary WD. For typical accretion velocities on the  order of thousands of km/s, the Reynolds number from the accretion flow onto the WD on a characteristic driving length scale of  $\sim 100$ km is of order $10^{16}$ \citep {nandkumarpethick84}. Similar considerations apply to a near-Chandrasekhar mass WD progenitor in the single-degenerate channel, in which the buoyant flame bubble also drives strong, high Reynolds-number turbulence \citep {niemeyerwoosley97}.

We assess the relative importance of nuclear burning and turbulence by computing the ratio of the specific  carbon specific nuclear burning rate to the turbulent dissipation rate at densities $\rho \simeq 10^7$ g cm$^{-3}$ at which a detonation is thought likely to arise \citep {woosley07}. The mean specific turbulent dissipation rate is 
%
$\epsilon_{\rm turb} = 10^{17} \left( {v_0 / 10^3\ {\rm km/s}}\right)^3 \left ({100\ {\rm km} / L}\right) \ {\rm erg\ s^{-1}\ g^{-1} }$.
%
Here $v_0$ is the RMS turbulent velocity on the integral length scale $L$. Additionally, the specific nuclear burning rate for carbon is highly sensitive to temperature,
%
$\epsilon_{\rm nuc} = 3.65 \times 10^{40} \rho_7^{2.5} X_{12}^2 \exp ({-65.894 / T_9^{1/3} })\ {\rm erg\ s^{-1}\ g^{-1} }$,
%
including electron screening \citep {garciasenzwoosley95}. Here $X_{12}$ is the mass fractional abundance of $^{12}$C, $\rho_7$ is the mass density in units of 10$^7$ g cm$^{-3}$, and $T_9$ is the temperature in units of 10$^9$ K. 

\begin{figure*}[h]
	\begin{center}
		\includegraphics[width=1.5\columnwidth]{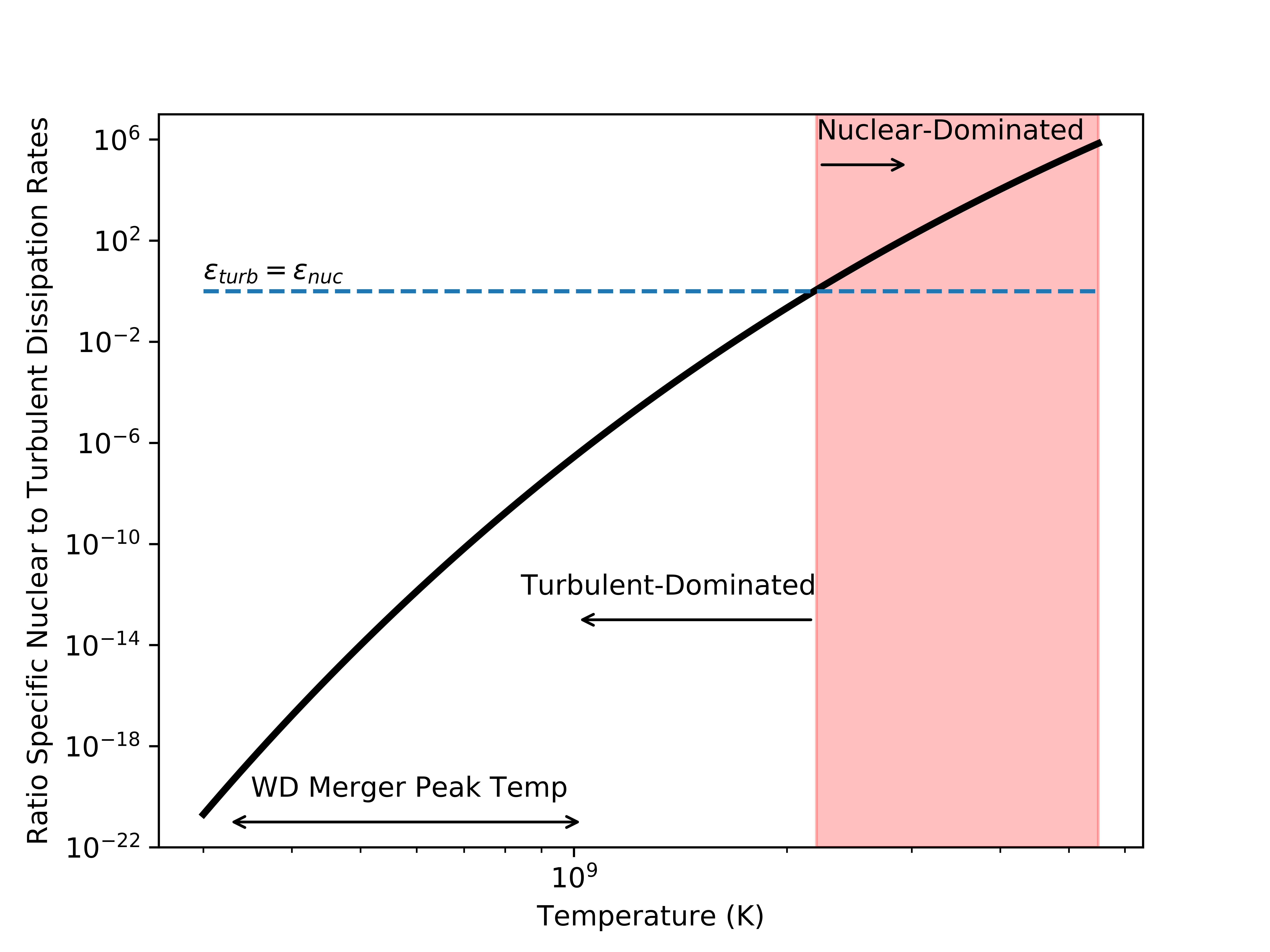}
		\caption{Plot of the ratio of the specific nuclear generation rate $\epsilon_{\rm nuc}$ to the specific turbulent dissipation rate $\epsilon_{\rm turb}$, as a function of temperature.  The thick solid line shows $\epsilon_{\rm turb} / \epsilon_{\rm nuc}$ as a function of temperature, for the representative values $v_0 = 2 \times 10^3$ km/s, $L = 100$ km, $\rho_7 = 1$, and $X_{12} = 0.5$. The horizontal dashed line demarcates the equality of the specific nuclear burning and turbulent dissipation rates, $\epsilon_{\rm turb} = \epsilon_{\rm nuc}$. The range of peak temperatures achieved in simulations of binary carbon-oxygen WD mergers from \citet {danetal14}  is  shown at bottom.}
		\label {fig:enucratio}
	\end{center}
\end{figure*}

We plot the ratio $\epsilon_{\rm nuc} / \epsilon_{\rm turb}$ versus temperature in figure \ref {fig:enucratio}. This plot demonstrates that {\it heating in carbon-oxygen WD mergers is initially energetically dominated by turbulent dissipation by up to 20 orders of magnitude}. It is also apparent from this figure that an increase of temperature by up to an order of magnitude is required for the fuel in merging carbon-oxygen WDs to enter into the nuclear-dominated regime. The energetics points towards turbulence as playing the dominant role in this process.

The dominant role of turbulent dissipation  has important ramifications for the physical properties of the burning at ignition.  At a critical temperature ($T \simeq 7 \times 10^8$ K at $\rho = 10^7$ g cm$^{-3}$), carbon burning overcomes neutrino losses. The relative significance of turbulence upon the burning surface can be quantified by the dimensionless Karlovitz number, Ka $= \sqrt {(u^3 / s_l^3) l / L }$, where $u$ is the RMS velocity on the integral scale $L$, and $s_l$ and $l$ are the laminar flame speed and thickness, respectively \citep {aspdenetal08}. When Ka $< 1$, turbulence plays a minor role on the scale of the flame, and the flame remains laminar. For large Ka $\gg 1$,  the flame is disrupted by the turbulence, and exists in the distributed burning regime. In the distributed burning regime, turbulent mixing strongly dominates the electron conduction, and  the products of nuclear burning are isobarically turbulently mixed, with thermal and species diffusion playing negligible roles \citep {aspdenetal08}. At densities $\rho \simeq 10^7$ g cm$^{-3}$, the characteristic laminar flame speed is $s_l \sim 4 \times 10^3$ cm/s, and the flame thickness $l = $ 4 cm  \citep {timmeswoosley92}. For a typical WD merger driven on length scales $L \simeq 100$ km with RMS turbulent velocity $u = 1 - 2 \times 10^3$ km/s, the Karlovitz number Ka $ \simeq 10^3 - 10^4$. Consequently, in typical carbon-oxygen WD mergers, nuclear burning is initiated in the distributed burning regime, in strong contrast to the single-degenerate scenario of Chandrasekhar-mass WDs, where nuclear burning is initiated as a flamelet \citep {niemeyerwoosley97}.  {\it Crucially, the  central assumption  of the Zel'dovich gradient mechanism, namely that detonation initiation arises through the development of a laminar flame on a static background,  does not apply to nuclear burning in the highly turbulent distributed burning regime. In particular, the Zel'dovich gradient mechanism omits the dominant role turbulence plays in both heating and transport in this regime.} 


Similar criticisms of the Zel'dovich gradient mechanism have been described previously -- e.g. \citet {niemeyer99}. However, the Zel'dovich mechanism is still frequently invoked in the literature. We proceed to describe the physics of the turbulently-driven detonation mechanism.

\section {Turbulently-Driven Detonation Mechanism}

In the turbulently-driven detonation mechanism, carbon-rich electron-degenerate fuel is heated by strong turbulent dissipation, initiating the formation of a hot spot,  and subsequently leading to supersonic burning and a detonation. The pressure support of the fuel is primarily derived from electron degeneracy, and as a result, the fuel does not rapidly respond to the turbulent heating. Turbulent dissipation is intrinsically highly intermittent, and is distributed inhomogeneously throughout the burning volume \citep {kolmogorov62, oboukhov62}.   In particular, in subsonic turbulence, the turbulent cascade is locally dissipated in highly-intermittent vortical structures, consistent with a log-Poisson statistical distribution of turbulent dissipation \citep {sheleveque94, dubrelle94}. This intermittent dissipation gives rise  to an intermittent temperature distribution on small scales, and naturally preconditions the carbon-rich fuel for a detonation. Furthermore, while the fuel must be at sufficiently low density to be in the distributed burning regime, turbulence suffices to both precondition and ignite the fuel. A strong pre-existing deflagration front is therefore not a prerequisite to achieve a detonation, as in previous deflagration-to-detonation transition (DDT) models \citep {khokhlovetal97}.

Hot spots arise from local temperature maxima in the turbulent flow. These hots spots  have a nuclear burning timescale $\tau_{\rm nuc} =  {e_{\rm int} / \epsilon_{\rm nuc} }$ required to increase the specific internal energy $e_{\rm int}$. Additionally, each turbulent hot spot of size $r$ is associated with an eddy turnover time, $\tau_{\rm edd} = r / v_r$, where $v_r$ is the Kolmogorov turbulent RMS velocity on the length scale $r$, $v (r) = v_0 \left(r / L  \right)^{1/3}$.

Because carbon burning is  dominated by turbulent dissipation in the distributed burning regime, in most local temperature maxima, the eddy turnover time is shorter than the nuclear burning timescale. Under these conditions, the hot spot will turbulently mix into the surrounding cooler material, and dissipate more rapidly than nuclear burning can increase its temperature. However, for temperature fluctuations at the  upper tail of the temperature distribution, the eddy turnover time may be longer than the nuclear burning timescale, at which point the eddy will undergo a nuclear runaway. One-dimensional models of distributed burning estimate a critical minimum scale of order $\simeq 1$ km for the onset of supersonic burning and detonation initiation \citep {woosley07}. For scales $r \simeq 1$ km, and for typical turbulent driving parameters $v_0 = 2 \times 10^3$ km/s and $L = 100$ km,  the nuclear burning timescale  falls beneath the eddy  turnover time  $\tau_{\rm nuc} < \tau_{\rm edd}$ at a critical temperature $T_{\rm crit} \simeq 2 \times 10^9$ K. Moreover, because of the  temperature sensitivity of the carbon nuclear burning rate, a large change in any of the turbulent driving parameters results in only a small change in the critical temperature $T_{\rm crit}$, which is generally expected to closely correspond to the transition into the nuclear-heating dominated regime. 

The possible importance of turbulent dissipation in initiating a detonation front in the distributed burning regime was first suggested by \citet {woosley07}. Pioneering 3D simulations of the distributed burning regime have been  subsequently carried out by \citet {aspdenetal08} and \citet {fennplewa17}. However, to date, no simulation in the literature has been carried into the detonation phase, motivating the current investigation.





\section {Simulation Methodology} In order to understand the physics of the turbulently-driven detonation initiation mechanism, we have carried out 3D simulations. We employ the FLASH hydrodynamics code, and utilize the split piecewise parabolic method (PPM) hydrodynamics solver.  We use an equation of state which includes contributions from nuclei, electrons, and blackbody photons, and which supports an arbitrary degree of degeneracy and special relativity for the electronic contribution \citep{Timmes_2000}. Nuclear burning is incorporated using a 19-isotope network with 78 rates described by \cite {weaveretal78}, and optimized in a hard-wired implementation detailed by \cite {timmes99}.


The fully-periodic domain is chosen to have size $L$ = 100 km, with an initially uniform  mass density $10^7$ g cm$^{-3}$, temperature $10^8$ K, and zero velocity. The initial composition is assumed to be equal proportions of carbon and oxygen. The momentum of the simulation is then driven to a steady state using a large-scale stochastic forcing routine which has been extensively verified against other numerical simulations, and validated against experiments \citep {benzietal08, arneodoetal08, benzietal10}. The stirring is done over waveumbers 1 - 4 with a smooth  paraboloidal injection of power, a solenoidal weight $\xi = 0.5$ providing roughly equal power in compressible and incompressible modes, and an autocorrelation time of 0.05 s \citep {federrathetal10}. Nuclear burning is turned on once the model achieves steady-state turbulent velocity statistics, establishing the zero point of our simulation clock, $t = 0.0$, at which point the models had mean temperatures $T_{\rm mean} \simeq 10^9$ K. Four simulations were conducted, with spatial resolutions varying of $64^3$, $128^3$, $256^3$, and $512^3$. Each simulation had identical turbulent driving strengths, achieving 3D RMS velocities $v_{\rm rms} = 2.2 \times 10^3$ km/s in steady-state, with the highest resolution simulations having 3\% higher turbulent RMS velocities than our lowest-resolution model. Under these conditions, the mean Karlovitz number Ka $= 8 \times 10^3$, and nuclear burning in accurately treated in a fully distributed fashion without incorporating electron conduction or a flame model.  We additionally verified isotropy and the low-order scaling exponents of the turbulent velocity statistics to within a few percent of accepted values  using the second- and third-order longitudinal velocity structure functions.


\begin {table}
\begin{ruledtabular}
\begin{tabular}{llll}
\textrm{Resolution}&
\textrm{$T_{\rm mean}$ (K)}&
\textrm{$t_{\rm det}$ (ms)}\\
\colrule
$64^3$  & $1.12 \times 10^9$ & 12  \\
$128^3$ & $1.17 \times 10^9$ & 14 \\
$256^3$ & $1.17 \times 10^9$ & 13  \\
$512^3$ & $1.18 \times 10^9$ & 15 \\
\end{tabular}
\caption {A summary of key results for the runs in this paper, including the mass-weighted mean temperature and time at detonation.}
\end{ruledtabular}
\end{table}

\section {Simulation Results} 
\begin{figure*}[h]
	\begin{center}
		\includegraphics[width=1.9\columnwidth]{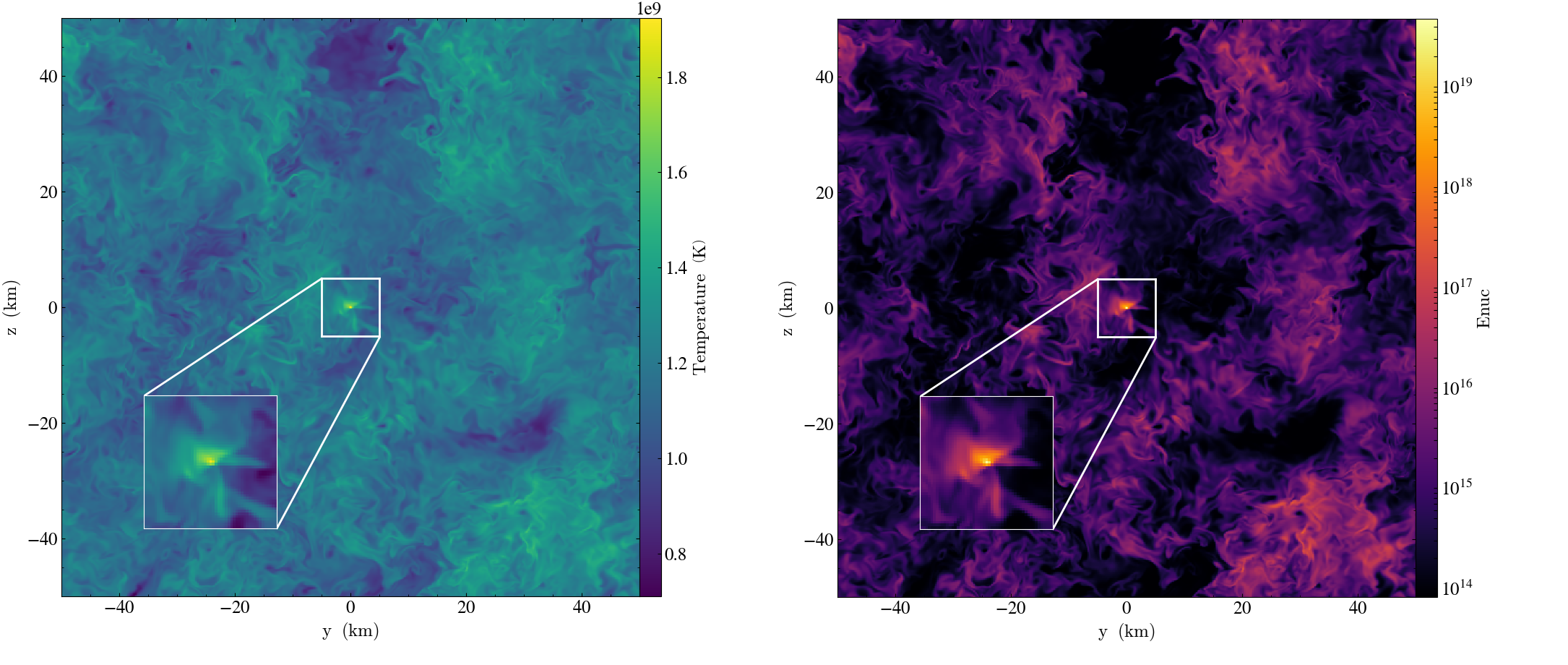}
		\caption{Slice plots of temperature and specific nuclear energy generation rate in the $y$-$z$ plane for the $512^3$ run. The point of maximum temperature is centered in the middle of these slices. The slices are taken at $t = 14$ ms, just prior to the onset of detonation for the $512^3$ run. The insets show zooms around the point of maximum temperature and nuclear energy burning, which subsequently develops into a detonation.}
		\label {fig:temp_enuc}
	\end{center}
\end{figure*}

We examine the results of 3D numerical simulations of strongly-driven turbulent distributed nuclear burning. We first examine the spatial distribution of temperature and nuclear burning. In figure \ref {fig:temp_enuc}, we show slice plots in the $y$-$z$ plane of both temperature and log specific nuclear burning rate for the highest-resolution $512^3$ run. These slices are centered about the maximum temperature, and taken a time just prior to the onset of  rapid nuclear burning and detonation initiation. Because of the temperature sensitivity of the carbon burning rate, the burning rate is strongly correlated with the temperature distribution.   The insets in figure \ref {fig:temp_enuc} show the resolved hot spot, of order a few km in extent, which develops into a detonation. 



\begin{figure*}[h]
	\begin{center}
		\includegraphics[width=1.8\columnwidth]{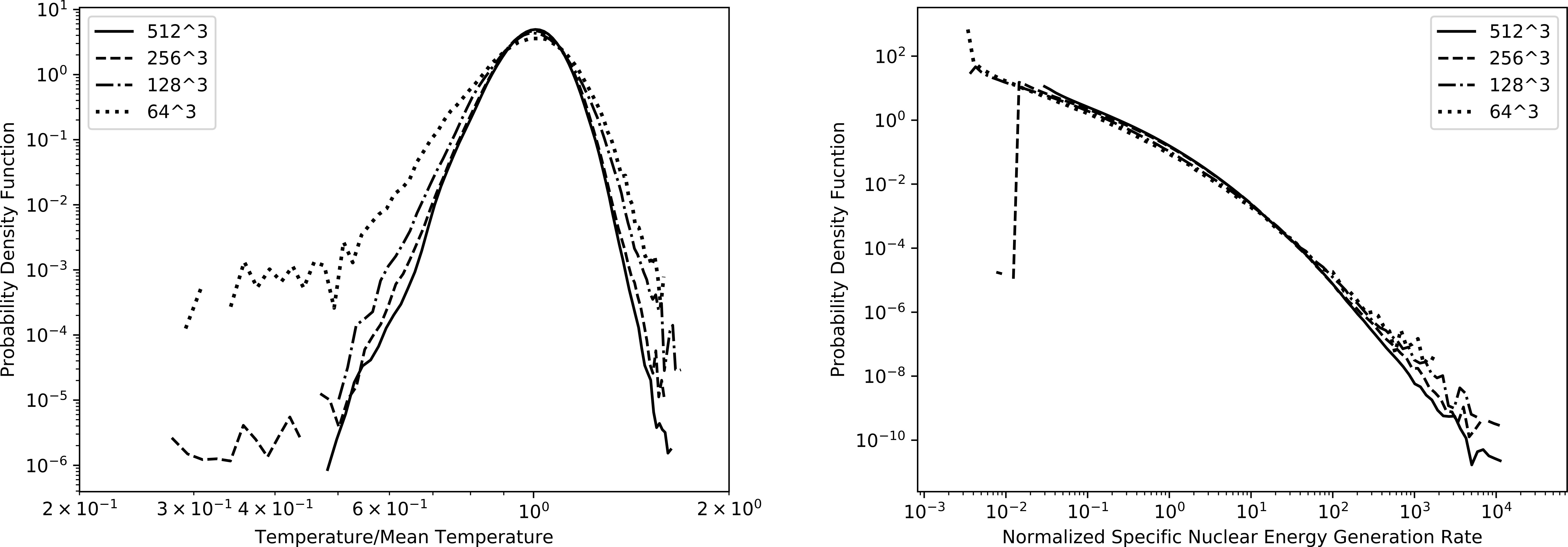}
		\caption{The PDFs for the temperature and  the specific nuclear burning rate, for each of the four runs, $64^3$, $128^3$, $256^3$, and $512^3$. Each PDF is shown just prior to the onset of detonation for that model. In both plots, the PDF is normalized to the mean value of each model.}
		\label {fig:temp_enuc.pdf}
	\end{center}
\end{figure*}

Turbulent fluctuations in the dissipation rate give rise to a statistical distribution of temperatures and also of nuclear burning rates. We quantify this temperature distribution in the one-point probability distribution functions (PDFs) of temperature and specific nuclear burning rates, shown in figure \ref {fig:temp_enuc.pdf}, for each of our four runs. The temperature PDF is approximately Gaussian, in agreement with experiments of grid-generated wind tunnel turbulence, which find Gaussian temperature PDFs in the absence of mean temperature gradients \citep {jayeshwarhaft91}. 


Additionally, we quantify the turbulent enhancement of the carbon burning rate in figure \ref {fig:temp_enuc.pdf}, where we plot the PDF of the specific carbon burning rate, normalized to the mean rate, for all runs. Because the carbon burning rate is highly sensitive to temperature, the relatively modest fluctuations in the temperature result in enhancements in the carbon burning rate by four orders of magnitude above the mean.

\begin{figure*}[h]
	\begin{center}
		\includegraphics[width=2.0\columnwidth]{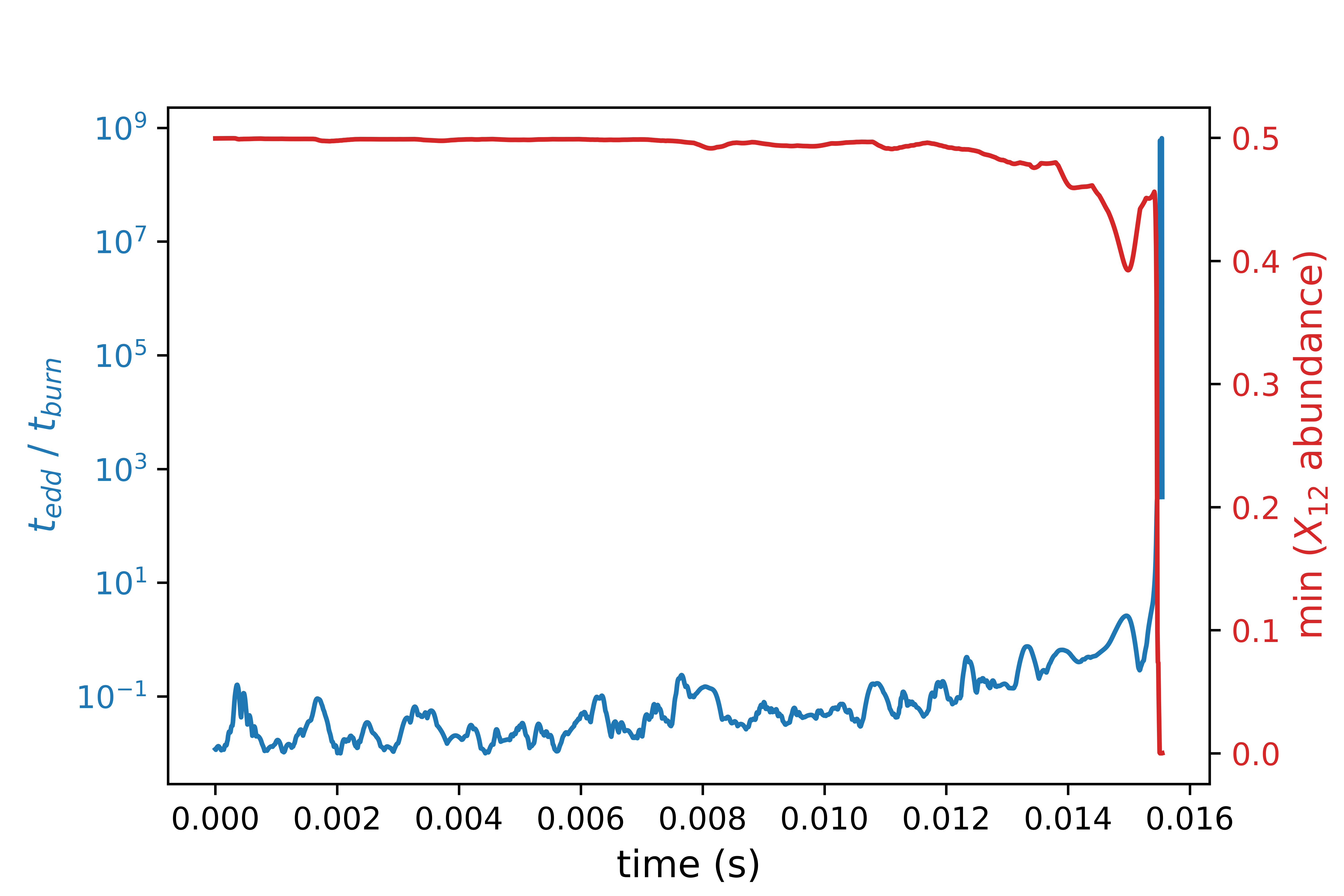}
		\caption{Plot of the ratio of the eddy-turnover timescale versus the carbon nuclear burning timescale (left axis, in blue), and the minimum $^{12}$C abundance (right axis, in red) versus simulation time. Local hotspots develop stable nuclear burning before turbulently mixing into the background on a $t_{\rm edd} \simeq 1$ ms timescale. At $t = 0.015$ s, a detonation develops centered at the hot spot of figure  \ref {fig:temp_enuc}. }
		\label {fig:c12vstime}
	\end{center}
\end{figure*}

In figure \ref {fig:c12vstime}, we show the time-evolution of the ratio of the eddy-turnover time $t_{\rm edd}$ to the burning timescale $t_{\rm burn} = X(^{12}$C) / $\dot {X}(^{12}$C) at the temperature maximum,  as well as the minimum $^{12}$C abundance at the location of the temperature maximum for the $512^3$ model. Initially, this ratio $t_{\rm edd} / t_{\rm burn}$ in the 3D simulation is already orders of magnitude larger than that predicted by the mean temperatures in the absence of turbulence as shown in figure \ref{fig:temp_enuc}, because of the enhancement in the nuclear burning rate due to intermittent temperature fluctuations. The ratio  $t_{\rm edd} / t_{\rm burn}$  is still initially less than unity, meaning that the nuclear burning is initially stable. As the simulation evolves, the ratio increases periodically as hot spots develop and are turbulently mixed into the background. At $t = 0.015$ s, or approximately three global eddy turnover times, a hot spot develops in which nuclear burning develops rapidly, with $t_{\rm burn} / t_{\rm edd} \simeq 10^9$. At this point, the nuclear burning develops supersonically, and a detonation is initiated. 


A key question is to what extent the simulation results depend upon the spatial resolution; that is, whether the simulations are spatially and temporally converged. At the lower resolution models of $64^3$ and $128^3$, the temperature PDF exhibits extended tails in comparison to those at higher resolutions. The resulting simulations detonate at slightly earlier times and lower mean temperatures than higher-resolution models.  In contrast, the models at $256^3$ and $512^3$ demonstrate a high degree of convergence. In particular, the mean temperature at the time of detonation is converged to within 1\% at the highest resolution.

Previous work on 3D simulations of SNe Ia has emphasized the importance of temporally resolving thermonuclear burning \citep {Hawley_etal_2012, kashyapetal15}. The temporal resolution criterion may be characterized by the maximum fractional change  $C_{\rm burn}$ of internal energy imposed per timestep, $\Delta t_{\rm burn} \leq \min (C_{\rm burn} e_{\rm int} / \epsilon_{\rm nuc})$. However, because the spatial resolution in these local simulations is orders of magnitude finer than global SNe Ia simulations, the CFL timestep alone implies a stringent burning timestep. In particular,
 our $512^3$ model has $C_{\rm burn} \simeq 0.1$.  In comparison, the most stringent burning timestep achieved in global 3D SNe Ia simulations is  $C_{\rm burn} = 0.2 - 0.3$ \citep {Hawley_etal_2012, kashyapetal15}. Convergence tests conducted by imposing an even smaller $C_{\rm burn}$ with one-half the value implied by the CFL did not produce any substantive differences from the models presented here.


\section {Conclusion} Previous physically-motivated detonations in the carbon-oxygen WD core in 3D double degenerate WD simulations, e.g. \citep {pakmoretal10, kashyapetal15}, have adopted criteria based upon the combined density and temperature within a given cell or smoothed particle hydrodynamics (SPH) particle to infer  whether a detonation is likely according to the Zel'dovich gradient mechanism. 
The resulting simulations are typically too slowly-declining \citep {pakmoretal10}, with too great a viewing angle dependence \citep {molletal14}, too much polarization \citep {bullaetal16}, and predict too few events \citep{liuetal18}, in comparison to normal SNe Ia. This disagreement between simulations and observations poses questions about the viability of the double-degenerate channel as a primary explosion channel for the majority of SNe Ia, and has motivated the introduction of alternative models, such as colliding WDs \citep {kushniretal13}. 

However, the imposition of the Zel'dovich gradient mechanism in global SNe Ia simulations neglects the important role of turbulence on detonation initiation. Turbulently-driven detonation initiation within the distributed burning regime naturally leads to a wider range of mean density and temperature conditions than previously realized. This broader range of predicted detonation initiation conditions has ramifications for leading SNe Ia channels, in which the onset of the detonation is predicted to arise under highly dynamical and turbulent conditions.

{\bf Acknowledgements}  RTF gratefully acknowledges conversations with Enrique Garc{\'i}a-Berro,  a close collaborator who tragically passed away in 2017, which inspired this work. RTF also thanks the Institute for Theory and Computation at the Harvard-Smithsonian Center for Astrophysics for visiting support during which a portion of this work was undertaken. RTF acknowledges support from NASA 17-ATP17-0115. This work used the Extreme Science and Engineering Discovery Environment (XSEDE) Stampede 2 supercomputer at the University of Texas at Austin's Texas Advanced Computing Center  through allocation TG-AST100038, supported by National Science Foundation grant number ACI-1548562 \citep {townsetal14}. 

\software
We utilize the adaptive mesh refinement code FLASH 4.0.1, developed by the DOE NNSA-ASC OASCR Flash Center at the University of Chicago. For plotting and analysis, we have made use of yt \citep {Turk_2011}, \url {http://yt-project.org/}.


\end {document}